\documentclass[10pt, conference]{IEEEtran}
\IEEEoverridecommandlockouts
\usepackage{cite}
\usepackage{amsmath,amssymb,amsfonts}
\usepackage{algorithmic}
\usepackage{graphicx}
\usepackage{textcomp}
\usepackage{xcolor}
\usepackage{makecell}
\usepackage{colortbl}
\usepackage{multirow}
\usepackage{enumitem}
\usepackage{mwe}
\usepackage{changepage}
\usepackage{subcaption}
\usepackage{cleveref}
\emergencystretch=\maxdimen
\usepackage{array}
\def\BibTeX{{\rm B\kern-.05em{\sc i\kern-.025em b}\kern-.08em
    T\kern-.1667em\lower.7ex\hbox{E}\kern-.125emX}}

\begin{document}

\title{Are you cloud-certified? Preparing Computing Undergraduates for Cloud Certification with Experiential Learning \\
}
\author{\IEEEauthorblockN{Eng Lieh Ouh}
\IEEEauthorblockA{\textit{School of Computing and Information Systems} \\
\textit{Singapore Management University}\\
Singapore \\
elouh@smu.edu.sg}
\and
\IEEEauthorblockN{Benjamin Kok Siew Gan}
\IEEEauthorblockA{\textit{School of Computing and Information Systems} \\
\textit{Singapore Management University}\\
Singapore \\
benjamingan@smu.edu.sg}
}

\maketitle

\begin{abstract}
Cloud Computing skills have been increasing in demand. Many software engineers are learning these skills and taking cloud certification examinations to be job competitive. Preparing undergraduates to be cloud-certified remains challenging as cloud computing is a relatively new topic in the computing curriculum, and many of these certifications require working experience. In this paper, we report our experiences designing a course with experiential learning to prepare our computing undergraduates to take the cloud certification. We adopt a university project-based experiential learning framework to engage industry partners who provide project requirements for students to develop cloud solutions and an experiential risk learning model to design the course contents. We prepare these students to take on the Amazon Web Services Solution Architect - Associate (AWS-SAA) while doing the course. We do this over 3 semester terms and report our findings before and after our design with experiential learning. We are motivated by the students' average 93\% passing rates over the terms. Even when the certification is taken out of the graded components, we still see an encouraging 89\% participation rate. The quantitative feedback shows increased ratings across the survey questions compared to before experiential learning. We acknowledge concerns about the students' heavy workload and increased administrative efforts for the faculty members. We summarise our approach with actionable weekly topics, activities and takeaways. We hope this experience report can help other educators design cloud computing content and certifications for computing students in software engineering.
\end{abstract} 


\section{Introduction}
Cloud Computing adoption has seen significant growth over the last decade. Cloud computing is a distributed computing paradigm that enables access to virtualized resources, including computers, networks, storage, development platforms or applications. These resources can be unilaterally requested, provisioned and configured by the user with minimal interaction with the cloud provider. The growth in cloud computing presents challenges for companies to source cloud expertise to support their business, particularly small and medium-sized enterprises with limited resources. While cloud technologies have evolved at a significant pace, the development of the computing curriculum in the further and higher education sector training has lagged \cite{foster2018cloud}. Organizations are losing revenue due to a cloud expertise deficiency, and a shortage of cloud expertise is holding businesses back \cite{TheCostOfCloudExpertise}. It can be assumed that the deficiency in cloud skills can at least partially be attributed to a lack of skilled graduates coming through the employment pipeline.

Professional certifications provide a parallel path to classroom learning for aspiring software engineers. Students will obtain a transferable credential to compete in the job market by achieving certification from one of the major cloud technology organisations. The industry white paper \textit{Cloud Skills and Organisational Influence: How Cloud Skills Are Accelerating the Careers of IT Professionals} \cite{Anderson2017WhitePC} provides evidence that candidates with professional certifications related to cloud development and operations have dramatically more influence over their organisation’s adoption and expansion of cloud services than IT professionals without relevant certifications.

One challenge for undergraduates who aspire to be software engineers to prepare for these cloud certifications is their lack of working experience in cloud computing. Many of these cloud certifications recommend a certain degree of working experience to be ready to take the examination. For example, the Amazon Web Services (AWS) Solution Architect - Associate (AWS-SAA) recommends at least 1 year of hands-on work experience designing cloud solutions that use AWS services \cite{AWSSAA}. However, work experience requirements should not discourage institutes of higher learning from modeling courses after certification materials, nor should it discourage current students from taking certification exams \cite{knapp2017maintaining}. Experiential learning can close the gap between the abstract concepts taught in the classroom and the skills needed for students to succeed once they join the workplace \cite{holmes2018dimensions}. Experiential learning offers the possibility of learning connected to work and professional endeavour and the possibility of ideas being developed through that which can be observed or experienced \cite{reynolds2009wild}.

Rapid advances in cloud computing and related technologies have made incorporating cloud computing into an existing curriculum challenging. One core reason for the challenges is the extent of cloud-related concepts and the plethora of choices to implement among cloud tools and technologies. The classical instructional approach to start from a rough set of curricular content and student feedback after each semester does not work for cloud curricula because each semester typically brings multiple changes in the field. These changes may be due to competition among cloud providers, improvements in design and implementation, additional offerings, or retiring specific product offerings. Traditional lecture content may be less successful for cloud concepts. Instead, students should be encouraged to explore and experiment through practical activities in the form of experiential learning.

Experiential learning is the process whereby knowledge is created through experience transformation. SMU-X project-based experiential learning framework \cite{SMU-X} provides opportunities for students to take on real-world challenges by collaborating on projects with corporate, non-profit and government organizations. On the other hand, Kolb \cite{kolb2014experiential} offers a systematic statement of the theory of experiential learning and its modern applications to education, work, and adult development. Kolb models the underlying structures of the learning process based on the latest insights in psychology, philosophy, and physiology. In our earlier work, we adapt and propose Kolb's experiential learning model with risk management process \cite{lieh2018exploring} for course design. In this paper, we combine the experiential risk learning (ERL) model and SMU-X framework to design this course. The ERL model provides the critical design steps to develop the course materials for experiential learning, while the SMU-X framework provides the key project activities for a project-based experiential learning journey.

Many education conference working groups are researching teaching cloud fundamentals and involving certifications in the curriculum. Foster et al. \cite{foster2018cloud} identified that there is limited current literature on cloud adoption in education curriculum and using cloud platform resources. Paterson et al. explore ways to help educators incorporate cloud computing into their courses and curricula by mapping industry job skills to knowledge areas and learning objectives \cite{paterson2021designing}. This paper adds value to these discussions on teaching cloud architecture designs and incorporating certifications within the course. We use identified knowledge areas and learning objectives in our mapping to the learning objectives of AWS-SAA certification.

In this paper, we explore project-based experiential learning SMU-X \cite{SMU-X} and experiential risk learning (ERL) model  \cite{lieh2018exploring} to prepare our CS undergraduates to take on the AWS-SAA certification. We describe existing literature on cloud certifications for undergraduate computing students. Section III describes the process of designing cloud computing and certification in a course. Section IV describes our course design changes, and Section V evaluates the outcomes of implementing the proposed course design. We summarise the key takeaways in Section VI, threats to validity in Section VII and the conclusion in Section VIII.

\section{Literature Review}
Teaching cloud computing has been relatively new in undergraduate computing programs. Certifying students in this area is even newer. In this section, we review the existing literature on teaching cloud concepts to undergraduates and preparing them for cloud certification.

Sommerville \cite{sommerville2013teaching} discusses the teaching of cloud computing from a software engineering perspective in terms of Sensitization (Telling students about something and how it is used), Practice (students are given some tuition in the practical elements of a topic) and Principles (abstract the fundamental principles of a topic and present these to students). The author expressed that we have reached the stage of sensitization and believe that something on this topic should be included in all courses. For Practice, the value of using IaaS in a software engineering course is that it is technically very simple to support server-based project work, which without the cloud, may have required access to dedicated hardware. Teaching about scale has been a perennial problem in software engineering. The author felt that the cloud provides a vehicle for experimentation with the technical issues of very large-scale systems. 

Eickholt and Shrestha \cite{eickholt2017teaching}  presented options for teaching Cloud Computing while using a physical cluster. It allows for the quick adoption of new tools and concepts. It does not penalize students for additional study or experimentation as does the pay-as-you-go model associated with cloud providers. They also acknowledge that Some effort on the part of faculty and some flexibility from the institution will likely be needed to implement a physical cluster to support teaching Cloud Computing. 

Deb, Fuad and Irwin \cite{deb2019module} argue that, for substantial coverage of cloud computing concepts and skills, the relevant topics need to be integrated into multiple core courses across the undergraduate CS  curriculum rather than creating additional standalone modules. They present their implementation of three such modules and our classroom experiences while deploying them. The assessment results clearly show that the students could relate to the topics very well, found them to be  interesting enough to explore and retain, and developed significant interest and confidence after the interventions.

Working groups convened multiple times at an computer science education conference \cite{paterson2021planning,paterson2021designing,foster2019toward} to explore ways to help educators incorporate cloud computing into their courses and curricula by mapping industry job skills to knowledge areas (KAs).  They identified, organized, and grouped student learning objectives (LOs) and developed these KAs and LOs in a repository of learning materials and course exemplars.

Podeschi and Debo \cite{podeschi2022integrating} discuss integrating cloud computing and AWS Cloud Practitioner Certification into a Systems Administration Course for computing undergraduates. The authors use labs, quizzes and final projects to prepare for the certification exam. Being the first iteration of incorporating cloud technology into the curriculum, only 5 students took the examination, and 3 passed. There is also a working group looking into the role that certifications play in the cloud computing curriculum. Their study framed in terms of the stakeholders and the connections between students, graduates, institutions (colleges/universities), vendors, other creators of courses and certifications and employers. Their findings \cite{paterson2022motivation} conclude that there is significant value for students and employers in the inclusion of certifications as an aspect of industry-relevant courses in cloud computing. One of their recommendations is to include certification exams within the academic credit for courses as this will motivate students to achieve certifications.

This paper extends the existing knowledge in teaching cloud computing, focusing on modifying an existing course design for solution architecture with cloud computing concepts and preparing students to take on the AWS-SAA certification with experiential learning.

\section{Designing Course For Cloud Computing and Cloud Certification }
This section details our course design process to teach cloud computing and prepare our students to be cloud certified with experiential learning. It involves
\begin{itemize}
    \item[A.] Identify Cloud Computing Certification
    \item[B.] Mapping Course Objectives for Certification
    \item[C.] Application of Experiential Risk Learning model
    \item[D.] Application of SMU-X framework
    \item[E.] Managing the Cloud Certification
\end{itemize}

\subsection{Identify Cloud Computing Certification}
The right cloud computing platform selection dictates the degree of support we can expect from the cloud provider and the certification to be provided to the students. There are many established cloud providers (e.g., Amazon Web Services, Microsoft Azure, Google Cloud). Their certifications differ in terms of acceptance in the industry. We adopt an approach \cite{ouh2021integration} to analyse the industry demand for the type of certifications required and survey the certification demand by crawling job postings targeting the roles of a software developer, engineer or architect. The approach uses text analytics to automatically parse the postings and derive the set of high demand information technology (IT) certifications in the job postings.

\subsection{Mapping Course Objectives for Certification}
Most cloud providers have their certification learning materials tailored to their cloud technologies. To use these materials in an undergraduate course, a faculty can use the cloud provider's materials as-is or merge them with other materials that teach non-vendor-specific concepts. Using the cloud provider's materials as-is decouples the need to review and update the materials when the cloud provider's materials change. If there is an existing course, there is a need to evaluate the fit to the cloud provider's topics. As a rule of thumb, the learning objectives of the current course should at least 75\% fit the learning objectives of the certification.

\subsection{Application of Experiential Risk Learning model}
The Experiential Risk Learning model provides the key design steps to develop the course materials. Experiential Risk Learning (ERL) model \cite{lieh2018exploring} presents a cyclical learning model through four stages that encompass risk management concepts. 

The Concrete Experience (CE) stage comprises the learner observing risk scenarios whereby independent events can happen to an existing system, resulting in a non-trivial negative impact. These scenarios should be repeatable to ensure that it does not happen by chance. The Reflective Observation (RO) stage involves the learner attempting to evaluate the severity of these risks. During this stage, the learner can have more guidance from the instructor or additional lessons on underpinning knowledge leading to this compromise. This underpinning knowledge is the essential technical or non-technical information related to the event generated, how this event can access the system, or how the system works. With this knowledge, the learner is expected to prioritize the items in a risk scenario log. During the Abstract Conceptualization (AC) stage, the learner needs to model a solution to mitigate the prioritized risks. The solution model must effectively mitigate the same risk from happening and should be clearly defined, implementable and testable. The solution model might introduce more risk scenarios, which can be added to the risk scenario log but are not required to be addressed in the same cycle. The Active Experimentation (AE) stage involves the learner implementing the solution and executing it to mitigate the prioritized risks. The cycle repeats with the learner entering the concrete experience stage again but with the updated risk scenario log from the earlier cycle. The learner can attempt to address other risks in another cycle.

\begin{figure}
  \centering \includegraphics[width=180pt]{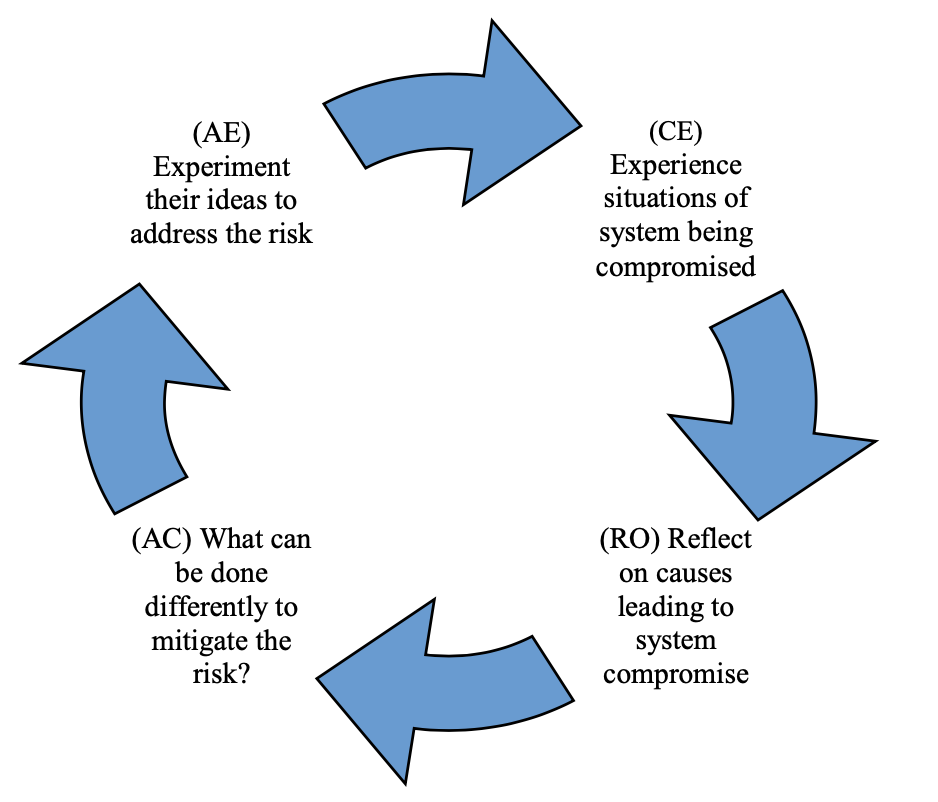}
  \caption{Experiential Learning Model}
  \label{experiential_learning}
\end{figure}

\subsection{Application of SMU-X framework}
 Many certification exams prefer candidates with working experiences on their platform or technology. Applying experiential learning closes this gap. SMU-X is a project-based framework that combines academic with experiential learning through the heavy use of projects from industry partners. SMU-X challenges students to use their knowledge and skills to tackle real-world problems through interdisciplinary approaches and activities. Industry partners and faculty are involved in active mentoring so that students benefit most from this collaborative relationship; SMU-X students get a better understanding of what it means to use theory learnt outside the classroom; faculty learn how real-world adapts theories; industry partners deepen their learning through faculty and students’ findings \cite{pan2015smu}\cite{pan2017industry}.

The framework comprises 5 key process activities - Challenge Identification (CI), Kick-off Briefing (KB), Learning and Application (LA), Solution Building (SB) and Presentation Day (PD) as shown in Figure \ref{experiential_learning}. The CI activity allows the faculty members and project champions to identify and scope project challenges appropriate for the course. The KB activity allows the project champions to present their project challenges to students and can involve on-site visits. The LA activity is when faculty members introduce new theories, concepts and tools for students to apply. Students at this stage can interact with project champions to further understand the problem. Students during this period will reflect and brainstorm ideas on how to solve the problem and add value to the business. Students can also discuss the matter with faculty members on the solutions. The SB activity involves the students researching and prototyping the solution, which may include on-site visits and deployments. Students can also receive critiques from their project champions and faculty members to further develop the solutions. The PD activity allows students to present their recommendations to their project champions. During this final stage, project champions will provide feedback to students, and faculty members will grade the students.

\subsection{Managing the Cloud Certification}
There are administrative efforts to manage the students in taking these certifications. The students need to plan the taking and timely submission of the certification results for end-of-term grading. Students who wish to retake certification examinations must understand the blackout period if failed for the first time and allocate sufficient time for retake and result submission. If the cloud provider provides discount vouchers, the faculty member must ensure the timely allocation of these vouchers to the students. If the course allows further subsidy either at the school level or other subsidy schemes, the faculty member has to track the timely submission of these claims. This is crucial to avoid rejecting late claims by the claims department. For example, claims after more than 2 months of fee payment are not accepted.

\section{Proposed Course Design}
This section details the considerations and rationale we make to design our course. Table \ref{tab:table-actionableweeks} shows an example of the weekly topics and activities for the proposed course design.

\subsection{Identify Cloud Computing Certification - AWS SAA}
Based on our approach to evaluating the job demands for technology certifications, we decided on the Amazon Web Services Solution Architect - Associate (AWS SAA) certification, one of the highly demanded certifications in these job postings. The full details can be found in our earlier paper \cite{ouh2021integration}.

\subsection{Mapping Course Objectives for Certification - ITSA}
We decided to teach designing solutions on the cloud using an existing course on IT solution architecture (ITSA). We reviewed and mapped the ITSA learning objectives against proposed cloud knowledge areas and learning objectives in earlier works \cite{foster2018cloud} and the learning objectives of the AWS-SAA \cite{AWSSAExamGuide}. We find a good fit whereby ITSA focuses on software architecture design with an emphasis on performance, availability and security qualities that align well with AWS-SAA, which focuses on a strong understanding of the AWS Well-Architected Framework that also includes performance, availability and security. Table \ref{tab:table-mapLOs} shows the mapping between AWS-SAA learning objectives against course learning objectives based on existing work \cite{foster2018cloud} and the last column shows the experiential learning activities based on the SMU-X framework and Experiential Risk Learning (ERL) model, which are described in detail in the following sections.

In a typical 15-week term (including a recess week and two weeks of final examination) for ITSA, we use 2 teaching weeks to introduce the architectural thinking concepts and project requirements and another 4 weeks on architectural quality attributes in security, availability and performance. We use the next 3 weeks after recess to cover advanced topics such as continuous integration and deployment while allowing the teams to develop the project solution. The last 2 weeks are for project preparation and presentations. A sample of the weekly ITSA topics and activities is also shown in Table \ref{tab:table-actionableweeks}.

\subsection{Application of Experiential Risk Learning model - ITSA}
This section describes how we design the availability, security and performance activities as shown in Table \ref{tab:table-mapLOs} with regard to the experiential learning of Concrete Experience (CE), Reflective Observation (RO), Abstract Conceptualization (AC) and Active Experimentation (AE). 

For each quality, we apply the four stages of experiential learning for the learners to experience the qualities in action, reflect, conceptualize and experiment for further insights. These sessions follow a similar format where the students will form groups and experience situations where the quality is compromised. With the observations, they reflect on the root causes leading to the system being compromised and the impact on the system. They subsequently decide what they can do differently to improve the quality and mitigate the risks. The next activity allows them to experiment with their ideas and evaluate whether they can mitigate the risks identified. This cycle of experiential learning for each quality iterates; after one cycle, additional potential new risks can be introduced. The following scenarios are described with an initial configuration that involves the students setting up a three-tiered web architecture with a java proxy server, a Tomcat web application and a MySQL database.

\subsubsection{Analyzing Designs for High Availability}
For the CE activity, the students experience the potential loss of web application availability. For example, killing the software process of the web application or unplugging the network connection between the proxy server and the web application. The students can and should discover many other scenarios that can cause the loss of availability of the web application. For example, any single point of failure in the architecture design can cause a loss of availability, such as the down of the proxy server or web application due to coding bugs or device issues. For the RO activity, the students reflect on the symptoms and single points of failures affecting the system's availability. For example, how did the proxy server handle the exception? Did the proxy server retry after failure? The students can discuss within their group and the instructor to understand the symptoms and identify potential root causes of the risks. For the AC activity, the students conceptualize what can be done differently to mitigate the key risk. For example, how to design for redundancy and clustering of the web application to mitigate this risk? There can be many mitigation measures, and the students have to discuss and draw sequence and deployment diagrams to illustrate their designs. Possible design considerations include horizontal or vertical scaling of the web application, how to check for failures and the failover mechanism of the web application instances. For the AE activity, the students can implement their idea to improve the availability and quality of the system. For example, the students can implement and test a redundant and clustered web application configuration. The proxy server can be programmed to detect failure and failover to the redundant web application instance. The exercises are modified to include AWS technologies to prepare the students to take the AWS-SAA certification examination. For example, using AWS Elastic Load Balancer (ELB) to distribute the traffic from failed AWS Elastic Compute Cloud (EC2) instances to other working instances for availability.

\subsubsection{Analyzing Designs for Security}
For the CE activity, the students experience the potential security compromise of the web application. For example, sniffing using OWASP Zap \cite{zap} exposes the login credentials contained within unencrypted network traffic. For the RO activity, the students reflect on the potential threats and vulnerabilities of the system. For example, what are the root causes of the proxy server or the web application leading to this compromise? What are the other threats and vulnerabilities concerning this two-tier web application setup? For the AC activity, the students can conceptualize what can be done differently. For example, how to design for security such that the data are secure in transit, storage, and archive. They can draw sequence and network diagrams and design firewall rules to illustrate their designs. For the AE activity, the students can implement their idea to improve the security quality of the system. For example, the students can create a digital certificate and implement a secure socket layer (SSL) for the web application. The security exercises are modified to include AWS technologies. For example, AWS Access Control Lists (ACLs) and Security Groups (SGs) are firewalls to allow or deny traffic entering the subnets and the instances respectively for security.

\subsubsection{Analyzing Designs for Performance}
For the CE activity, the students experience the initial performance of the web application. For example, a performance test using Apache JMeter \cite{jmeter} for a given set of concurrent users can help to create a baseline of the web performance. For the RO activity, the students reflect on their performance trends and identify potential bottlenecks of the system. For example, what are the CPU, memory and network utilization levels on devices hosting the proxy server and web application? Where is the bottleneck? For the AC activity, the students conceptualize what can be done differently. For example, how can they tune to improve performance by addressing the bottleneck identified earlier? Will implementing a cache helps? They can illustrate their design using sequence diagrams or table calculations. For the AE activity, the students can implement their idea to improve the performance quality of the system. For example, if there is a bottleneck but the memory is underutilized at 40\%, can the memory allocated to the web application be increased to improve performance? These performance exercises are modified to include AWS technologies. For example, using AWS Auto-Scaling and Amazon Machine Images (AMIs) to spawn more EC2 instances to meet increased workload demand.

\subsection{Application of SMU-X framework - ITSA}
Based on the SMU-X framework, we engage industry partners as project champions with a mutually agreed arrangement for them to engage with our students throughout the semester. The project champions are involved in providing project requirements, evaluating the student's proposal and final implementation. The project case studies differ across terms, and industry partners can propose more than one case study. There are at least 2  case studies, and the students can choose one of the proposed case studies to work on. They can seek clarification from the mentor throughout the semester.

For the challenge identification stage as shown in the second and third columns of Table \ref{tab:table-actionableweeks}, the faculty will discuss with the mentor before the start of the course to align their expectations against the course objectives and the certification requirements. For example, the proposed solutions to the project case studies must be deployed on the AWS platform, and requirements must also be aligned with the quality attributes. These discussions usually take weeks and up to 2 months, and the deliverable includes a project statement that describes the project requirements and any supporting documents or data. For example, most case studies involve integrating with external systems. The project champions must prepare the dataset and the external APIs to support the student teams during their implementation. These documents must be prepared early so that the students are not hindered by the delay of these deliverables during the semester.

The kick-off briefing takes place in the second week of the term to introduce the project champions to the students and to allow the students to understand the project challenges early. The solution building and learning \& application stages take place from week 3 till week 12. Each team is required to submit a project proposal in week 7 to be graded by the faculty member and project champion. At the same time, the project champion can validate the students' understanding and give recommendations for areas to improve. The teams can improve on their solution till week 13, when they present the concrete implemented solution on the presentation day.

\renewcommand{\arraystretch}{1.4}
\begin{table*}
\caption{\label{tab:table-mapLOs} Mapping Certification Objectives to Knowledge Areas / Learning Objectives to Experiential Learning}
\centering
\begin{tabular}{| p{5.5cm} | p{5.5cm} | p{5.5cm} |}
\hline
\textbf{AWS SAA Learning Objective} & \textbf{Mapping to Knowledge Areas and Course Learning Objectives (subset)} & \textbf{Experiential Learning (subset)} \\
\hline
\multirowcell{8}{}{
\begin{itemize}[leftmargin=0.3cm,topsep=2pt,itemsep=0.5em, after=\vspace{\baselineskip}]
        \item Design secure access to AWS resources
        \item Design secure workloads and applications
        \item Determine appropriate data security controls
    \end{itemize} 
}
&   
\textbf{Cloud Security, Privacy, Policy and Ethics}
\begin{itemize}[leftmargin=0.3cm,topsep=2pt,itemsep=0.5em, after=\vspace{\baselineskip}]
        \item Demonstrate the use of security controls in the cloud computing environment
        \item Design a secure architecture model using security design principles
        \item Understand security threats in cloud computing Test for threat vulnerabilities in a cloud computing application system
    \end{itemize} 
& 
\textbf{Experiential Risk Learning Model  (Security)}
\begin{itemize}[leftmargin=0.3cm,topsep=2pt,itemsep=0.5em, after=\vspace{\baselineskip}]
        \item Implement AWS ACLs and SGs for network security
    \end{itemize}

\textbf{SMU-X framework - Solution Building}
\begin{itemize}[leftmargin=0.3cm,topsep=2pt,itemsep=0.5em, after=\vspace{\baselineskip}]
        \item Implement to fulfil the security requirements of the SMU-X project
    \end{itemize} 
    
\\        
\hline
\multirowcell{8}{}{
\begin{itemize}[leftmargin=0.3cm,topsep=2pt,itemsep=0.5em]
        \item Design scalable and loosely coupled architectures
        \item Design highly available and/or fault-tolerant architectures
    \end{itemize} 
}
&   
\textbf{Fault Tolerance, Resilience and Reliability}
\begin{itemize}[leftmargin=0.3cm,topsep=2pt,itemsep=0.5em, after=\vspace{\baselineskip}]
        \item Configure and deploy a fault tolerant cloud service solution that meets the minimum SLA requirements
        \item Define the concept of service availability, and classify different levels of availability
    \end{itemize} 
& 
\textbf{Experiential Risk Learning Model (Availability)}
\begin{itemize}[leftmargin=0.3cm,topsep=2pt,itemsep=0.5em, after=\vspace{\baselineskip}]
        \item Implement cross-region failover using AWS Route53
    \end{itemize}

\textbf{SMU-X framework - Solution Building}
\begin{itemize}[leftmargin=0.3cm,topsep=2pt,itemsep=0.5em, after=\vspace{\baselineskip}]
        \item Implement to fulfil the availability requirements of the SMU-X project
    \end{itemize} 
\\        
\hline   
\multirowcell{8}{}{
\begin{itemize}[leftmargin=0.3cm,topsep=2pt,itemsep=1em]
        \item Determine high-performing and/or scalable storage solutions
        \item Design high-performing and elastic compute solutions
        \item Determine high-performing database solutions
        \item Determine high-performing and/or scalable network architectures
    \end{itemize} 
}
&
\textbf{Cloud Elasticity and Scalability}
\begin{itemize}[leftmargin=0.3cm,topsep=2pt,itemsep=0.5em, after=\vspace{\baselineskip}]
        \item Describe the vertical and horizontal scaling approaches that can be deployed to enhance the performance of a cloud service
        \item Configure and deploy a cloud service solution that addresses the requirements of horizontal or vertical performance scaling
        \item Understand load balancing mechanisms on the middleware
    \end{itemize} 
& 
\textbf{Experiential Risk Learning Model (Performance)}
\begin{itemize}[leftmargin=0.3cm,topsep=2pt,itemsep=0.5em, after=\vspace{\baselineskip}]
        \item Implement the AWS Auto-Scaling to handle increased workload
        \item Implement the AWS Elastic LoadBalancer to distribute traffic across EC2 instances
    \end{itemize}

\textbf{SMU-X framework - Solution Building}
\begin{itemize}[leftmargin=0.3cm,noitemsep,topsep=2pt, after=\vspace{\baselineskip}]
        \item Implement to fulfil the performance requirements of the SMU-X project
    \end{itemize} 
\\        
\hline    
\multirowcell{8}{}{
\begin{itemize}[leftmargin=0.3cm,topsep=2pt,itemsep=1em]
        \item Design cost-optimized storage solutions
        \item Design cost-optimized compute solutions
        \item Design cost-optimized database solutions
        \item Design cost-optimized network architectures
    \end{itemize} 
}
&   
\textbf{Fundamental Cloud Concepts}
\begin{itemize}[leftmargin=0.3cm,topsep=2pt,itemsep=0.5em, after=\vspace{\baselineskip}]
        \item Explain the differences between leasing versus ownership of compute resources and compare the total cost of ownership  
        \item Discuss and critique the cost/performance characteristics and trade-offs of using scalable cloud services 
    \end{itemize} 
& 
\textbf{Experiential Risk Learning Model (Resource Cost)}
\begin{itemize}[leftmargin=0.3cm,topsep=2pt,itemsep=0.5em, after=\vspace{\baselineskip}]
        \item Use the AWS Cost Calculator to estimate resource costs
    \end{itemize} 

\textbf{SMU-X framework - Solution Building}
\begin{itemize}[leftmargin=0.3cm,topsep=2pt,itemsep=0.5em, after=\vspace{\baselineskip}]
        \item Implement to fulfil the budget requirements of the SMU-X project
    \end{itemize} 
\\        
\hline
\end{tabular}
\end{table*}

\begin{table*}
\caption{\label{tab:table-actionableweeks} Sample ITSA Weekly Topics and Activities}
\centering
\begin{tabular}{| p{2.2cm} | p{5cm} | p{5.8cm} | p{3.2cm} |}
\hline
\textbf{Week} & 
\textbf{Course Topic } &
\textbf{Process Activity} &
\textbf{Remarks} \\
\hline
\multirow{2}{*}{\makecell[l]{Before Week -8}} & 
\multirow{2}{*}{\makecell[l]{ }} & Identify Cloud Computing Certification & 
\multirow{2}{*}{\makecell[l]{Finalise Cloud Certification \\ and Course}}   \\
\cline{3-3}
& & Identify Course for Certification &\\
\hline
Week -8 to Week 1 & & SMU-X framework - Challenge Identification &  Finalise Course Project Requirements\\
\hline
Week 1 & Course Introduction &  &  \\
\hline
\multirow{2}{*}{\makecell[l]{Week 2}} & 
Introduction to Solution Architecture & SMU-X framework - Kick-Off Briefing & \\
\cline{2-2}
& Project Introduction & & \\
\hline

\multirow{2}{*}{\makecell[l]{Week 3}} & 
\multirow{2}{*}{\makecell[l]{Analyzing Designs for Resource Cost}} & SMU-X framework - Learning \& Application & \\
\cline{3-3}
& & Experiential Risk Learning Model &\\
\hline

\multirow{2}{*}{\makecell[l]{Week 4}} & 
\multirow{2}{*}{\makecell[l]{Analyzing Designs for Security}} & SMU-X framework - Learning \& Application & \\
\cline{3-3}
& & Experiential Risk Learning Model &\\
\hline

\multirow{2}{*}{\makecell[l]{Week 5}} & 
\multirow{2}{*}{\makecell[l]{Analyzing Designs for Availability}} & SMU-X framework - Learning \& Application & \\
\cline{3-3}
& & Experiential Risk Learning Model & \\
\hline
\multirow{2}{*}{\makecell[l]{Week 6}} & 
\multirow{2}{*}{\makecell[l]{Analyzing Designs for Performance}} & SMU-X framework - Learning \& Application & \\
\cline{3-3}
& & Experiential Risk Learning Model &\\
\hline
Week 7 & Project Proposal Presentation & SMU-X framework - Solution Building & Students submit Course Project Proposal \\
\hline
Week 8 & Recess Week & & Students book date for Certification Examination\\
\hline
Week 9 & Advanced Topics in Solution Architecture & SMU-X framework - Solution Building & Students take Certification Examination\\
\hline
Week 10 & Advanced Topics in Solution Architecture & SMU-X framework - Solution Building & Students take Certification Examination\\
\hline
Week 11 & Advanced Topics in Solution Architecture & SMU-X framework - Solution Building & Students take Certification Examination\\
\hline
Week 12 & Project Preparation & SMU-X framework - Solution Building & Students take Certification Examination\\
\hline
Week 13 & Project Presentation & SMU-X framework - Presentation Day & Students present Course Project Implementation\\
\hline
Week 14 & Preparation for Certification Examination & & Students take Certification Examination \\
\hline
Week 15 & & & Students submit results of Certification Examination\\
\hline
\end{tabular}
\end{table*}

\subsection{Managing the Cloud Certification}
For AWS certifications, AWS provides AWS Academy with online courses to support students learning and taking their certification. AWS Academy online course covers AWS technologies extensively with slides and practice labs that are sandboxed with usage restrictions. The labs gave the students hands-on experiences without the need to subscribe to an AWS account and provide personal and credit card information. AWS also provides a 50\% discount voucher for students to take their certification examination. The students are given the voucher in week 7 so that they can book their certification examination by week 8 and submit their results by week 15.

\section{Course Feedback and Analysis}
\subsection{Course Feedback Design}
This course is one of the core courses in a computer science (CS) program. CS students take this course in their year 3 of study. Students from the information systems (IS) program also take this course as their elective. The students are exposed to the two teaching methods - one without experiential learning from the academic year 2017-2018 Term 1 onwards for 5 terms (total of 324 students working on an internal project instead) and one with experiential learning from 2020-2021 Term 1 for another 3 terms (total of 133 students). We seek feedback using a survey from these groups of students on the effectiveness of the course with experiential learning compared to earlier terms without experiential learning. The survey is conducted anonymously before the term ends and only aggregated survey results are released to the instructor after the examination results are released.

The survey comprises metric-based questions rated on a 5-point Likert scale ranging from 1 – Strongly Disagree, 2 – Disagree, 3 – Neutral, 4 – Agree and 5 – Strongly Agree, focusing on the following dimensions. As we are integrating the cloud provider's materials, we want to understand from the students the overall learning experience in terms of the instructor's preparation, organisation, clarity and understandability and ability to stimulate interest in the subject. With experiential learning, timely and quality the feedback are essential for students to learn effectively. Lastly, we also want to know the overall rating of the instructor and the course.

The following sections analyse the survey results and further discuss the exam statistics after the students took the AWS-SAA examination. Besides the quantitative feedback, we also analyse the qualitative comments of the students.

\subsection{Analysis of Quantitative Feedback}
Table \ref{tab:table-quantiativefeedback} shows the survey results for the terms before and after applying experiential learning. 2019-2020T2 scored the highest but the student's enrollment is low at 11. To properly determine the trend, we calculate the average results taking into account the enrollment size before and after applying experiential learning. The average results for each question show an increasing trend after applying experiential learning. Initially, we are concerned that the inclusion of cloud providers' materials may be challenging for undergraduates to absorb, causing a downward trend in these ratings. However, the students can cope with the difficult degree of materials prepared primarily for working professionals. We are pleased with the slight increase in the ratings instead.

Table \ref{tab:table-awsstatistcs} shows the certification examination statistics when the students took the examination over the 3 terms when experiential learning is applied. The statistics are encouraging, averaging 93\% passing rates and an average score of 782 over 1000. They need 720 to pass. While the students in the first two terms (2020-21T1 and 2020-21T2) are required to take the certification examination as part of their grading, the students in the next term (2021-22T1) are not. It is motivating to know that at least 82\% of the students still took the examination with a 91\% passing rate.

\subsection{Analysis of Qualitative Feedback}
Many qualitative comments show that students understand the need for cloud computing skills and project-based experiential learning to improve their learning. Below is a curated set of comments.

\begin{adjustwidth}{0.5em}{0pt}
\emph{"I have a better understanding about ways to improve the performance, efficiency, reliability and security of a system hosted on AWS cloud services. I have also gained a better understanding of the various services offered by AWS."}\\
\emph{"The ability to design better architectures and a deeper understanding of AWS services and overall design patterns, development and much more. The opportunity to work with a real life sponsor and incorporating specific requirements into our project is also a valuable lesson."}\\
\emph{"ITSA has taught me a lot of useful knowledge on solution architectures and designs to keep a system secure and maintainable. I have learnt a lot as well about AWS's cloud services, which will be a good foundation for exploring AWS and other cloud providers. The course has also taught us to prepare for uncertainty, and that though a system cannot be 100\% secure or available there are many strategies to improve these aspects of the system. I believe this is a very valuable course that will stay with me further on in my career."}\\
\emph{"The project challenged us in a technical way, which forced out of our comfort zones."}\\
\emph{"I think the project requirements are something quite applicable in the real world for us to tackle."}\\
\end{adjustwidth}

There are some comments on the increased workload that the faculty needs to be aware of.
\begin{adjustwidth}{0.5em}{0pt}
\emph{"Make some resources available so that prior to starting of semester, people can prepare themselves for this intensive course."}\\
\end{adjustwidth}
\vspace{-1em}
With project-based experiential learning, the delivery of the project champions is crucial to the student's learning effectiveness. We do receive both encouragement and areas for improvement.
\begin{adjustwidth}{0.5em}{0pt}
\emph{"I like how their expectations and project requirements are clearly stated, and its very thorough and well planned. Through this project implementation, we have managed to effectively learn AWS."}\\
\emph{"The feedback that was provided to us during mid term was constructive in helping us improve on our implementation."}\\
\emph{"They were open to our different ideas of implementing solutions to their problems and also provided sufficient data for us to test our solutions."}\\
\emph{"The project sponsor is very supportive during presentations. They do not heavily criticize the work done, instead, they understand we are students and they provide constructive feedback on what can be improved."}\\
\emph{"they are not that clear of what they want, this gives us more flexibility in the project but the downside will be that we will not be sure if what we are doing meets their objectives."}\\
\emph{"functional requirements were not clear and specific enough The datasets provided was good. Though some of it is not realistically usable i.e. data was partial and had no linkage."}\\
\end{adjustwidth}
\renewcommand{\arraystretch}{1.5}
\begin{table*}
\caption{\label{tab:table-quantiativefeedback} Quantitative Feedback}
\centering
\begin{tabular}{| p{6cm} | p{0.6cm} | p{0.6cm} | p{0.6cm} | p{0.6cm} | p{0.6cm} | p{0.6cm} | p{0.8cm} | p{0.8cm} | p{0.8cm} | p{0.8cm} |}
\hline
\multirow{2}{*}{\makecell[l]{\textbf{Survey Questions and Ratings / Semester Term}}} & 
\multicolumn{6}{c|}{\textbf{Before Experiential Learning}} &
\multicolumn{4}{c|}{\textbf{After Experiential Learning}}\\
\cline{2-11}
& \textbf{2017-18T2} & \textbf{2018-19T1} & \textbf{2018-19T2} & \textbf{2019-20T1} & \textbf{2019-20T2} & \textbf{Avg} &
\textbf{2020-21T1} & \textbf{2020-21T2} & \textbf{2021-22T2} & \textbf{Avg} \\
\hline
Instructor's Preparation and Organisation &  
6.500 & 6.429 & 6.481	& 6.744 & 6.818 & \textbf{6.594} & 6.756	& 6.629	& 6.667  & \textbf{6.684}\\  
\hline
Instructor's Clarity and Understandability &  
6.600	& 6.381	& 6.426	& 6.744	& 6.818 & \textbf{6.594} & 6.756	& 6.629	& 6.667 & \textbf{6.625} \\  
\hline
Instructor's Simulation of interest in content &  
6.500	& 6.349	& 6.407	& 6.744	& 6.818 & \textbf{6.564} & 6.683	& 6.543	& 6.583 & \textbf{6.603}\\  
\hline
Quality and frequency of feedback &  
6.375 & 6.222 & 6.315 & 6.718 & 6.727 & \textbf{6.471} & 6.537 & 6.714 & 6.667 & \textbf{6.639}\\  
\hline
Overall Rating of the Instructor &  
6.538 & 6.437 & 6.481 & 6.744 & 6.909 & \textbf{6.622} & 6.805 & 6.714 & 6.667  & \textbf{6.729}\\  
\hline
Overall Rating of the Course &  
6.263 & 6.071 & 6.259 & 6.667 & 6.818 & \textbf{6.416} & 6.585 & 6.343 & 6.500  & \textbf{6.476}\\  
\hline
\end{tabular}
\end{table*}

\begin{table*}
\caption{\label{tab:table-awsstatistcs} AWS Certification Exam Statistics}
\centering
\begin{tabular}{| p{6cm} | p{0.6cm} | p{0.6cm} | p{0.6cm} | p{0.6cm} | p{0.6cm} | p{0.6cm} | p{0.8cm} | p{0.8cm} | p{0.8cm} | p{0.8cm} |}
\hline
\multirow{2}{*}{\makecell[l]{\textbf{AWS Certification Results / Semester Term}}} & 
\multicolumn{6}{c|}{\textbf{Before Experiential Learning}} &
\multicolumn{4}{c|}{\textbf{After Experiential Learning}}\\
\cline{2-11}
& \textbf{2017-18T2} & \textbf{2018-19T1} & \textbf{2018-19T2} & \textbf{2019-20T1} & \textbf{2019-20T2} & \textbf{Avg} &
\textbf{2020-21T1} & \textbf{2020-21T2} & \textbf{2021-22T2} & \textbf{Avg} \\
\hline
Number of Students &  
84 & 132 & 56 & 41 & 11 & 65 & 48 & 44 & 41 & 44\\  
\hline
Number of Students took AWS Certification Exam &  
N.A. & N.A. & N.A. & N.A. & N.A. & N.A. & 48 & 44 & 34 & 42\\  
\hline
Number of Students passed AWS Certification Exam &
N.A. & N.A. & N.A. & N.A. & N.A. & N.A. & 46 & 40 & 31 & 39 \\  
\hline
\% of Students passed AWS Certification Exam &
N.A. & N.A. & N.A. & N.A. & N.A. & N.A. & \textbf{96\%} & \textbf{91\%} & \textbf{91\%} & \textbf{93\%} \\  
\hline
Average Score for AWS Certification Exam (1000) &
N.A. & N.A. & N.A. & N.A. & N.A. & N.A. & \textbf{792} & \textbf{782} & \textbf{772} & \textbf{782}\\  
\hline
\end{tabular}
\end{table*}

\vspace{-1pt}
The students are given the opportunity to take the certification examination but not as a graded component in 2021-22 Term 2. Below are the comments when we survey why they proceed to take the certification exam.
\begin{adjustwidth}{0.5em}{0pt}
\emph{"To elevate my resume since I am going to start work soon in 1/2 years time."}\\
\emph{"I chose to take the certification examination because I feel like the certification will be a very useful thing to be displayed in my resume or LinkedIn and also I have taken the AWS CCP exam before so the learning curve is not as steep."}\\
\emph{"good to have on linkedin"}\\
\emph{"Force myself to study well to take the certification examination while still having to do well for the quizzes in case I fail the exam. If I pass the exam, its about 80\% which is also close to a grade A. Meaning no easy way out, good motivation to force myself to excel in both options."}\\
\end{adjustwidth}

Below are the students’ comments when we survey why they are not taking the certification exam.
\begin{adjustwidth}{0.5em}{0pt}
\emph{"I feel that the AWS certificate requires more time and commitment to cover materials not tested and I do not have bandwidth for additional tasks beyond what is absolutely necessary"}\\
\emph{"not very confident in the content that will be tested in AWS Certification Exam"}\\
\emph{"I chose not to take as I am currently interning part-time, and have insufficient time to revise the certification content outside of my school days."}
\end{adjustwidth}

\section{Key Reflections and Benefits Implementing the Proposed Course Design}
\subsection{Key Reflections}
\begin{itemize}
\setlength\itemsep{0.3em}

\item Selection of an appropriate cloud certification examination determines how the course can be modified. To achieve that, substantial efforts should be spent at the start to identify and, in our case, the job demands of these certifications.

\item Ensure a good fit of the course design learning objectives against the certification learning objectives. The similarity in evaluating the software qualities (availability, performance and security) allows us to quickly adapt the AWS materials to the proposed course design.

\item Identification of an appropriate project champion is critical to successfully applying the SMU-X framework. The project champion must understand the content's scope and the knowledge assessed for the certification. The project champion also has to provide timely support for the students to complete the project within the timeframe.

\item Teaching concepts and cloud computing materials provided by the cloud providers can be overwhelming for a course run in a typical 15 weeks term. We must balance teaching underlying concepts and practice exercises on cloud computing technologies. One strategy we use is give students materials to read up on the concepts first. During class, we discuss the concepts with examples and scenarios, followed by exercises and labs. 
\item There are two kinds of labs. One type is to illustrate the concepts, which requires students to draw diagrams or hands-on implementation locally on their computer. Another type is to apply the concepts of AWS technologies. Due to limited class time, we went for team-based and multiple labs simultaneously. Teams can decide for themselves. The teams are usually 5-6 members and max of 2 kinds of labs at the same time. Since the lab instructions can be carried out outside the class, students who want to do the other lab can still do it. We convene before the end of class to discuss what they learn from the labs.

\item There are additional efforts to include certification in a course. From the school and university perspective, external certification may not have sufficient academic rigour. For AWS SA - Associate certification, it comprises scenario-based multiple choices multiple-select questions. The examination has to be taken at an accredited testing location or virtually with an online proctor. To address the need for academic rigour, the certification accounts for less than 20\% of the student's final grade, and the faculty also sets additional AWS questions in the final examination. Some students have already obtained this certification. In such cases, their score is considered as long as they have taken the certification after matriculation. They are welcome to take other AWS associate certifications and use the score as part of grading too.
\item Faculty members have to be sufficiently competent to teach students for the certification. The faculty members teaching this course are also AWS-certified.
\item There are administrative efforts such as the examination logistics. Although the certification examination can be taken most of the time throughout the year, the students must complete the examination within the semester weeks for the marks to be accounted for in the final grade. AWS administration may take several business days to release the results, and this delay must be taken into account as the university has strict deadlines to complete grading. 
\item We do not moderate the certifications marks. The AWS certification is awarded based on 1000 marks, and we map it to the student's assessment marks accordingly. For example, if the score is 800 and accounts for 10\% of the total grade, the student is awarded 8 out of 10. We also do not consider if the student passes or fails the certification examination. The passing mark for AWS SAA is 720 out of 1000. Even if the student scores below 720, we will still map according to the student's assessment marks.
\item Another administrative effort for the faculty member is the handling of the examination fees. With the AWS discount voucher of 50\% for students, the student ends up paying USD75 for an attempt at the certification examination. This amount can be significant for certain students. We managed to arrange for funding both at the national and university level to cover these costs. At the national level, we have Industry Preparation for Pre-graduate (iPREP) Programme, which funds citizens to take certification courses and examinations if they pass the examinations. At the university level, we draw on funds for implementing experiential learning to support students who are not eligible for iPREP or have unfortunately failed the examination. This arrangement, including the claim process, requires substantial logistic efforts from the faculty and teaching team. However, it does provide a fair environment for students of diverse backgrounds to take this examination.
\item Most of the students see the value of obtaining the certification but not all. Some of them feel that the workload is too heavy with project-based experiential learning and taking the AWS certification. In the recent term, we experiment with a design whereby the students can choose to either take the certification or weekly quizzes as the grading component. We also ensure that the average and standard deviation of both certification and weekly quizzes are close, or else the students may feel one is easier than another. We see an encouraging 89\% participation rate of the AWS certification by the students. The downside to this design is that students who cannot choose between the two end up with even higher workloads by studying for both. We decide to revise this design to require the students to make their choice in the earlier weeks and focus on their choices for the rest of the term.
\end{itemize}
\vspace{-0.5em}
\subsection{Benefits for the students}
\begin{itemize}
\setlength\itemsep{0.1em}
\item Confidence in being able to design and deploy a quality solution architecture.
\item Expose students to explore AWS technologies and services. 
\item Minimize the cost of learning AWS services. Learning these services on their own may incur costs. They do not need to pay to use these AWS services for this course.
\item Receive subsidies and discounts to take a certification examination. AWS subsides the certification examination cost partially. Citizens can claim from iPREP and non-citizens can claim from a school funding.
\item Accredited skills with the AWS Solution Architect Associate (SAA) certification. 
\item Another benefit to the students is that they know more about the sponsor's companies, potentially leading to internship and job opportunities. 
\end{itemize}

\subsection{Benefits for the project champions}
\begin{itemize}
\setlength\itemsep{0.2em}
\item Accessibility to the students' deliverables for further evaluation and adoption. Sponsors commented that they see significant improvements in their outputs over the terms. All these prototypes are open source on GitHub, and the sponsors can access these repositories after the project ends. Understand and reuse ideas from the prototype to improve the software quality of their solutions in terms of performance, availability and security.
\item Possible interns or full-time employees. Many students are in their third or fourth (graduating) years, and the sponsors can tap into this pool. We have 3 interns and 5 full-time employees joining or have joined them. 
\item A win-win situation for the sponsor and the university. As a sponsor representative mentioned, "Our collaboration has been great for sponsors and the university. Through this, we've gotten not just sponsor's brand name out there but also the potential engineering experiences that our undergrads could have in the industry space globally.
\end{itemize}

\section{Threats to Validity}

This paper applies the proposed course design to the students taking ITSA for the AWS-SAA certification. Although the course is evaluated over many terms, we acknowledged that more studies need to be carried out for other cloud certifications.

The profiles of the students play a part in this study. We conduct this course for computing students taking information systems and computer science programs who are familiar with programming, design and business solutions. These students also taught using experiential learning and took the certification examinations during COVID-19 compared to those without experiential learning before COVID-19.

The application of this model depends on the instructors’ class conduct and interactions with the students. Although the class conduct for the three sections is based on the two instructors, there is still the potential threat of whether the study results will be valid for other instructors. We can only conclude that when we have other instructors teaching the same course using the proposed model.
\\
\section{Conclusion}
Cloud computing skills are increasing in demand for information technology professionals and graduates looking to start their careers in the IT industry. Preparing our undergraduates in Cloud Computing is a challenge given that it is a relatively new technology and covers multiple computer science disciplines, including networking, computing systems and software. 

Getting them certified is another challenge, given that these cloud certification examinations require working experience. This paper describes our experience preparing our undergraduates to be cloud certified. We identify the AWS Solution Architect - Associate (AWS-SAA) as the cloud certification based on a study of the demands in the job postings. We map the objectives of AWS-SAA to the knowledge areas and learning objectives of a cloud curriculum. We incorporate experiential learning activities based on our SMU-X learning framework and Experiential Risk Learning model.

The proposed course design implemented over 3 terms for 133 students shows that the students can take on these certifications with high passing rates of an average 93\%. The quantitative feedback shows increased ratings across the survey questions compared to before experiential learning. We acknowledge concerns about the students’ heavier workload and increased administrative efforts for the faculty members due to the inclusion of certifications. We describe the process to identify the cloud certification, map the learning objectives between the certification and our course and provide actionable weekly topics and activities. We hope this experience report can help other educators design cloud computing content and certifications for computing students in software engineering.




\bibliographystyle{IEEEtran}
\bibliography{references}
\end{document}